\documentclass[english,preprint]{aastex}
\usepackage{units}
\usepackage{graphicx}
\usepackage{amssymb}
\usepackage{esint}

\DeclareRobustCommand{\greektext}{%
 \fontencoding{LGR}\selectfont
 \def\encodingdefault{LGR}}
\DeclareRobustCommand{\textgreek}[1]{\leavevmode{\greektext #1}}
\DeclareFontEncoding{LGR}{}{}

\providecommand{\tabularnewline}{\\}

\slugcomment{}
\shorttitle{}
\shortauthors{}

\usepackage{babel}

\begin{document}

\title{Anomalous photon noise levels predicted for CMB measurements made
by the Planck satellite mission}

\author{Robin Booth\altaffilmark{1}}

\altaffiltext{1}{The Insular Institute, robin.booth@insularinstitute.org} 
\begin{abstract}
A fundamental assumption inherent in the standard \textgreek{L}CDM
Hot Big Bang (HBB) model is that photons lose energy as they are redshifted
due to the expansion of the universe. We show that for the Quasi-Static
Universe (QSU) model, in which photon energy is an invariant in the
cosmological reference frame, the photon number density in the universe
today is a factor of approximately 1600 less than in the standard
model. We examine some of the consequences for a number of processes
that occur during the thermal history of the early universe, including
primordial nucleosynthesis, the formation of neutral hydrogen (recombination),
and the evolution of the Cosmic Microwave Background (CMB) radiation.
We show that the QSU model predicts that the measured CMB photon noise
level will be a factor of $\sim40$ higher than the level that would
be observed assuming the standard HBB model. The CMB data that will
be collected by the recently launched Planck satellite mission provides
an ideal opportunity to test the validity of this prediction.
\end{abstract}

\keywords{cosmic microwave background, photon noise}

\section{Introduction}

\subsection{The Hot Big Bang Model}

The standard Hot Big Bang (HBB) model has formed the cornerstone of
cosmology for over 40 years, since the discovery of the $2.7K$ Cosmic
Microwave Background (CMB) by Penzias and Wilson in 1965. Over time,
the model has been extended and enhanced to provide a better fit with
the mass of data obtained from numerous ground and satellite based
astronomical observations. The main ingredients of the current so-called
{}`Concordance Model' include a hot Big Bang plus inflation, with
the addition of dark energy and cold dark matter to the mix. This
is often referred to as the Lambda + Cold Dark Matter (\textgreek{L}CDM)
model. One of the key features of the HBB model is the assumption
that the universe started off in a state of near infinite density
and temperature, and subsequently cooled as it expanded. This leads
to the concept of an early radiation dominated phase. As the universe
cooled it became possible for nucleons to condense out of the primordial
quark-gluon plasma. The standard model assumes that near equal amounts
of matter and anti-matter were formed at this stage. Subsequently,
these matter and anti-matter particles annihilated each other, releasing
energetic photons. It is these same photons that constitute the CMB
that we now observe. The small net imbalance between matter and anti-matter
resulted in the matter dominated universe that we now inhabit. Following
this matter generation epoch, the universe continued to expand and
cool until it became energetically favourable for nucleons to combine
to form heavier elements in a process of primordial nucleosynthesis. 

Further expansion reduced the temperature to a point at which previously
free electrons and protons were able to combine into neutral hydrogen.
As the temperature and matter density dropped further, a point was
reached at which the mean free path of CMB photons became comparable
to the size of the universe. At this stage, matter and radiation became
decoupled and fell out of thermal equilibrium and the CMB photons
were free to stream - the so-called Last Scattering Surface (LSS).
The CMB power spectrum that we observe today is therefore a snapshot
of the CMB at the LSS.

An important principle inherent in the standard model is that matter
density falls in proportion to $T^{3}$, whereas the energy density
of photons in the CMB radiation falls as $T^{4}$ because of the additional
energy loss as the photons are redshifted.

\section{The Quasi-Static Universe model}

The Quasi-Static Universe (QSU) is used here as a shorthand to refer
to the paradigm described in \citet{boo:2002}. The main feature of
this model is that the Planck scale is decoupled from the atomic scale
conventionally used as the as the basis for our measurement system.
The changes in the dynamical behaviour of the universe that result
from this modification can not only account for most of the problems
associated with the Big Bang model, but are also able provide a very
simple explanation for the apparent acceleration of the expansion
rate of the universe that has been observed in various high redshift
supernovae studies in recent years. One of the principal consequences
of the QSU paradigm is that the atomic scale, as defined by the de
Broglie wavelength of sub-atomic particles, is not an appropriate
reference frame for measuring gravitational phenomena or the behaviour
of photons. The QSU model is founded on the postulate that the correct
reference frame for these phenomena is in fact a cosmological frame
based on the size and energy content of the universe as a whole. In
such a frame, photons do not undergo any change in frequency, since
as far as they are concerned, the universe is static. Hence, it is
not meaningful to talk in terms of photons losing energy in this reference
frame. In transforming from the cosmological reference frame to our
conventional atomic frame, photons will be perceived to exhibit a
redshift as the scale factor of the universe with respect to the atomic
frame increases with time. The crucial difference between the QSU
model and the conventional formalism is that the relationship $E_{\gamma}=h\nu$
no longer holds true for photons emitted at times $t<t_{0}$, where
$t_{0}$ signifies the present time. The energy of such {}`old'
photons will remain constant, at the same value they possessed when
they were first emitted. However their power, i.e. the energy transferred
per unit time, will be reduced in proportion to their redshift, such
that $W=W_{0}.\nu\left/\nu_{0}\right.=W_{0}(z_{0}+1)$ , where $W_{0}$,
$\nu_{0}$ and $z_{0}$ are respectively the photon power, photon
frequency and cosmological redshift at the time of emission. Clearly,
such a modification to one the most fundamental equations in physics
will have a very significant impact on any phenomena that involve
redshifted photons, and in particular, the CMBR.

To understand the implications of the QSU model for the CMBR, we need
to review the way in which the standard Planck black-body distribution
law is applied. A summary of the standard derivation of this law is
provided in Appendix A. Present day observations of the CMB give a
value for the energy density of $U\simeq4\times10^{-14}Jm^{-3}$,
which from (\ref{eqn:U}) corresponds to a temperature of $T=2.7^{\circ}K$.
Conventionally, the next step is to take this temperature and use
equation (\ref{eqn:RhoGamma}) to calculate the photon number density,
giving a result of $N\simeq4\times10^{8}m^{-3}$. However, these formulae
are only valid for black-body radiation \emph{that is in equilibrium
with its surroundings}. It would be perfectly correct to use these
equations if we wished to deduce the photon number density for a black-body
with this temperature \emph{today}. In erroneously applying them in
the context of the relic CMBR generated by the Big Bang, we are perpetuating
the assumption inherent in going from (\ref{eqn:dU}) to (\ref{eqn:dN2})
- that photon energy is always equal to $h\nu$.

In the QSU model, red-shifted photons do not lose energy, so it is
not possible to determine the energy of an observed photon merely
by measuring its frequency. It is also necessary to know the thermal
history of the photon, i.e. its frequency when it was originally emitted.
It is this that determines the energy of the photon. In order to obtain
the correct result for the photon number density today that corresponds
to an observed energy density, we need to correct for the fact that
the photon energy $h\nu$ applied in going from (\ref{eqn:dU}) to
(\ref{eqn:dN2}) should reflect the energy at the time of emission,
or more precisely, its energy at the time it was in black-body equilibrium
with its surroundings.

We therefore need to apply a factor of $T_{obs}/T_{equi}$ to the
expression for photon number density in (\ref{eqn:dN2}), where of
$T_{obs}$ is the observed absolute temperature of the CMBR today,
and $T_{equi}$ is the temperature at the time of black-body equilibrium.
Clearly we do not know this figure with any precision. However we
can take an educated guess that is at least consistent with other
observational data and with plausible models for primordial nucleosynthesis.
One such model is the neutron decay variant of the Cold Big Bang (CBB),
which predicts a maximum reaction temperature of $\approx10^{10}K$,
with an equilibrium temperature of the order of $\approx10^{9}K$.
Applying the correction factor to equation (\ref{eqn:RhoGamma}) gives
a calculated photon number density of $\approx0.3m^{-3}$ - a factor
of $\sim10^{9}$ lower than for the standard Hot Big Bang (HBB) model.
This is very close to the measured baryon number density of the universe,
giving a photon-barion ratio of $\eta_{\gamma}\equiv n_{\gamma}/n_{B}\simeq1$.
It should be noted that this simplistic calculation is based on the
assumption that there are no intervening process that might change
the photon number density. As we shall see in the next section, this
assumption is unlikely to be valid.

\section{Implications for cosmological processes}

Having established that the initial value of $\eta_{\gamma}$ in the
QSU model is many orders of magnitude less than the conventionally
accepted value, the natural question to ask is: how does this affect
other cosmological processes? We now review the consequences of a
lower photon-barion ratio for three of the key stages in the thermal
history of the early universe:
\begin{enumerate}
\item Primordial nucleosynthesis
\item Recombination
\item CMB cooling
\end{enumerate}

\subsection{Primordial nucleosynthesis}

Arguably, one of the process most sensitive to changes in $\eta_{\gamma}$
is that of primordial nucleosynthesis. Applying the conventional value
of $\eta_{\gamma}\simeq2\times10^{9}$ to the standard model for HBB
nucleosynthesis results in predicted element abundances that are in
good accord with observational data. Relatively small changes in $\eta_{\gamma}$
will result in large changes to the predicted abundances, and would
therefore appear not to be compatible with observations. However,
it has been pointed out by \citet{agu:2001} that, provided that appropriate
changes are made to a range of initial parameters, including $\eta_{\gamma}$,
it is possible to construct alternative models for primordial nucleosynthesis
that will produce predicted element abundances that are still in accord
with observational data. One such model is the Cold Big Bang (CBB),
which takes a value of $\eta_{\gamma}\sim1$ as one of its initial
conditions. It is perhaps worth mentioning in passing that the neutron
decay variant of the CBB model predicts a photon energy of $0.78MeV$,
giving a baryon to photon energy ratio of $E_{\gamma}/E_{B}\simeq1200$.
Under the QSU scenario, this ratio should persist from the nucleosynthesis
epoch to the present day, and indeed, the observed value of $E_{\gamma}/E_{B}$
is very close to this value.

\subsection{Hydrogen ionization}

\begin{figure}[h]
\includegraphics[bb=50bp 460bp 550bp 780bp,clip,scale=0.8]{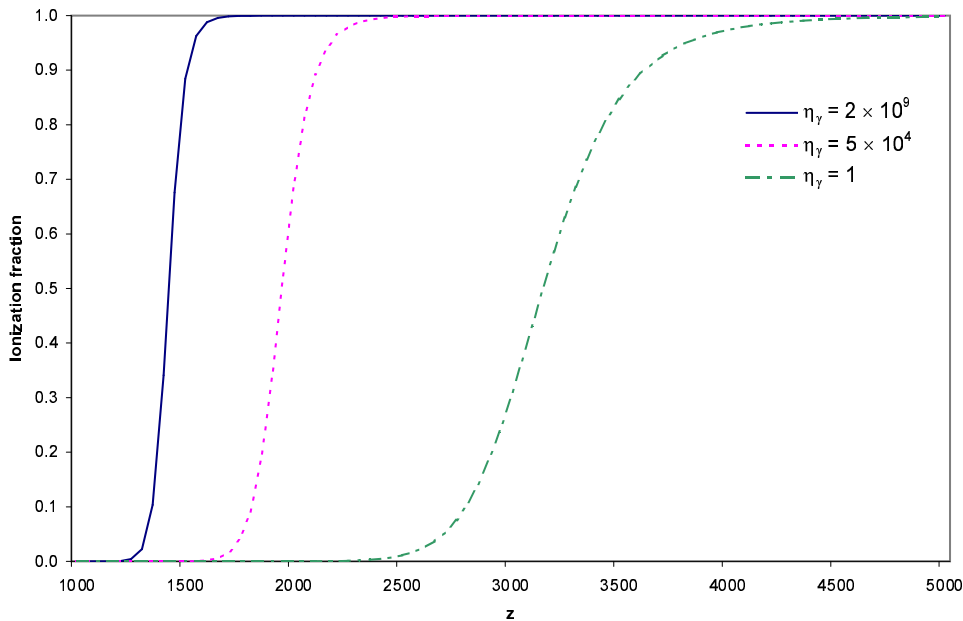}\caption{Ionization fraction as a function of redshift}
\label{fig:ionization}
\end{figure}

Another astrophysical measurement that is linked to $\eta_{\gamma}$
is the hydrogen ionization fraction. As the universe expands and cools,
protons combine with electrons to form neutral hydrogen when the energy
of CMB photons falls below the hydrogen ionization energy threshold.
The equilibrium ionization fraction $\chi_{e}$ as a function of temperature
is give by the Saha equation

\begin{equation}
\frac{{1-\chi_{e}}}{{\chi_{e}}}=\frac{{4\sqrt{2}\varsigma(3)}}{{\sqrt{\pi}\eta_{\gamma}}}\left({\frac{T}{{m_{e}}}}\right)^{\frac{3}{2}}e^{{B/T}}\label{eqn:saha}\end{equation}

from which it can be seen that the ionization fraction is also dependent
on $\eta_{\gamma}$. The ionization fraction can be expressed as a
function of redshift using the relation $T=2.73(1+z)K$. This is plotted
in Figure~\ref{fig:ionization} for a standard HBB cosmology with
$\eta_{\gamma}=2\times10^{9}$, and a CBB cosmology with $\eta_{\gamma}=1$.
This illustrates that a reduction in the photon to baryon ratio will
cause recombination to take place at a much higher redshift, with
the surface of last scattering for the CMB occurring at $z\simeq2600$,
as compared to $z\simeq1100$ for the standard model.

In practice, the situation is somewhat more complicated than that
depicted by the Saha equation because the system will not in fact
be in equilibrium. This is due to the fact that the process of electron-proton
recombination will itself release an energetic photon that will go
on to re-ionize another hydrogen atom. A more detailed analysis (see
\citet{pea:1999} for example) reveals that recombination proceeds
via a two-photon decay process, with the rate of re-ionization as
a function of redshift being given by

\[
\frac{d\ln\chi}{d\ln z}=60\chi z\frac{\Omega_{B}h^{2}}{\sqrt{\Omega h^{2}}}\]

where $\Omega_{B}$ is the barionic density parameter, $\Omega$ is
the total density parameter, and $h$ is the Hubble parameter.

The fact that the process of recombination results in the creation
of new photons is of particular relevance for the QSU model. In the
standard HBB model, the fact that the photon-barion ratio, at $\eta_{\gamma}=2\times10^{9}$,
is already very high before the beginning of the recombination era
means that the additional photons generated by the recombination process
will be of little consequence. Conversely, if one assumes an initially
low value for $\eta_{\gamma}=1$, such as might occur in the CBB scenario,
then the process of recombination will result in a large increase
in the photon number density. A value for $\eta_{\gamma}$ after the
recombination era can be estimated by taking the initial photon energy
for the CBB scenario, $0.78MeV$, and dividing it by the ionization
energy of the hydrogen atom, $13.6eV$. This gives a photon-barion
ratio of $\eta_{\gamma}\simeq5\times10^{4}$. The ionization fraction
as a function of redshift for this scenario is also plotted in Figure
\ref{fig:ionization}.

The Planck satellite will have sufficient sensitivity to measure the
angular power spectrum to a level of accuracy that will permit a more
accurate picture of the recombination era to be determined (\citet{esa:2005}).

\subsection{CMB measurements}

Measurements of the CMB spectrum can provide a wealth of information
about the early history of the universe (\citet{esa:2005}). The WMAP
satellite mission has already helped to refine the value of the main
parameters in the concordance model, including $\Omega$, $\Omega_{\Lambda}$,
and the Hubble constant. The recently launched Planck mission aims
to refine these measurements using the greater sensitivity of its
bolometric detectors to map the CMB at higher angular resolutions.
The satellite will also include polarization sensitive detectors,
enabling it to perform measurements of CMB polarization, which can
be used, amongst other things, to probe for gravitational waves that
might have been generated during an era of cosmic inflation. The sensitivity
of the detectors used on these satellite missions is determined largely
by the noise levels present in the detector chain. This includes thermal
noise for the detector chain itself, 1/f noise, Johnson noise, and
CMB photon counting noise. The performance of the bolometric detectors
on the Planck satellite has improved to the extent that sensitivity
is now limited primarily by the photon noise of the CMB (see appendix
\ref{sec:CMB-Photon-Noise}).

\begin{figure}[h]
\includegraphics[bb=50bp 350bp 550bp 780bp,clip,scale=0.8]{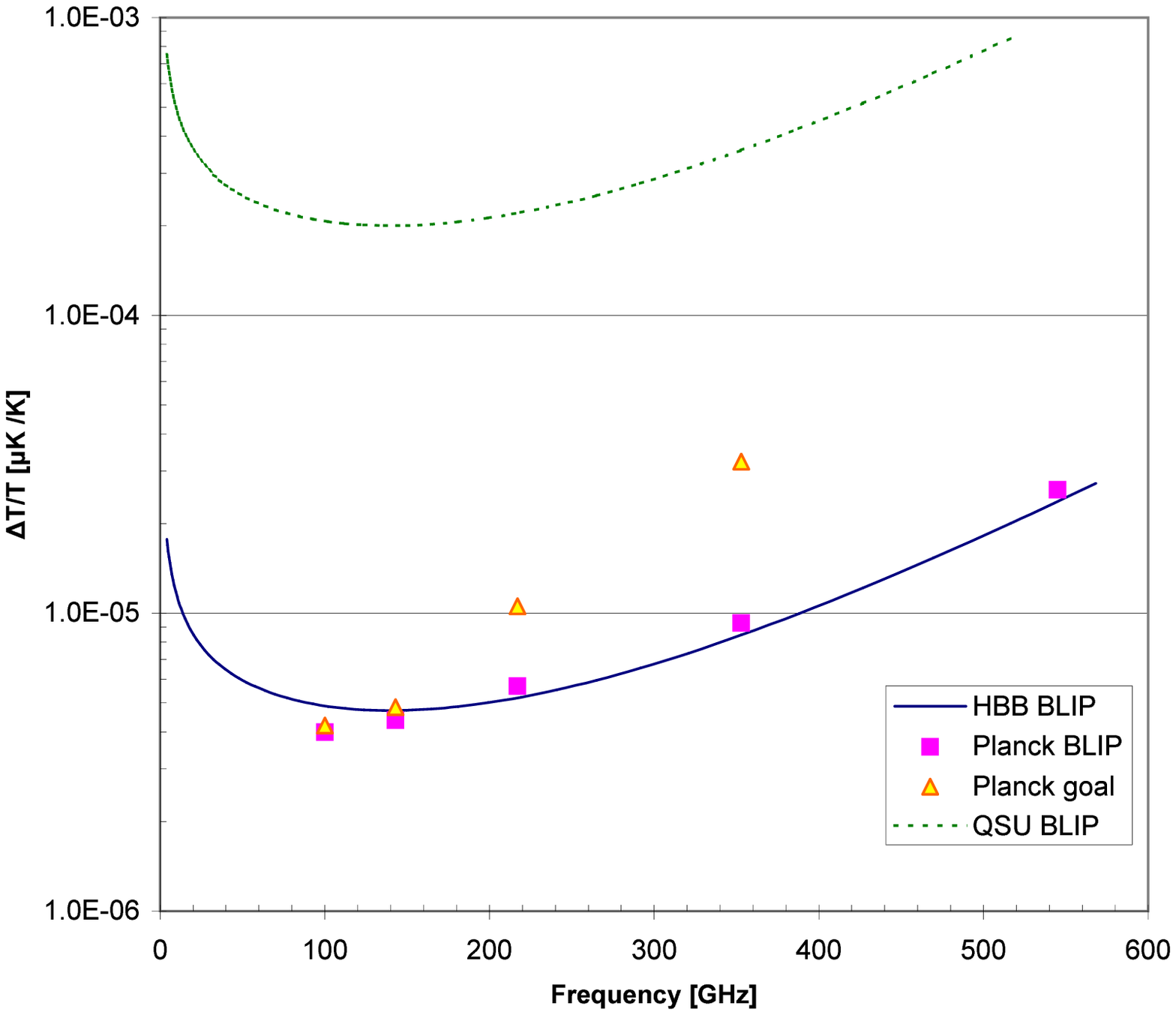}

\caption{CMB photon noise}
\label{fig:photon_noise}
\end{figure}

The performance goals for the Planck HFI are summarized in Table \ref{tab:Planck_HFI}.
The sensitivity figures represent the quadrature sum of the combined
sensitivities of all the bolometric detectors in that frequency band.
Figure \ref{fig:photon_noise} shows the sensitivities per pixel per
HFI detector, together with the theoretical CMB Background Limited
Infrared Photons (BLIP) according to the standard HBB model. From
this it can be seen that the noise performance of the most sensitive
detectors is at a level that is comparable to the CMB BLIP.

\begin{table}[h]
\begin{tabular}{|l|c|c|c|c|c|c|c|}
\hline 
Center frequency & GHz & 100 & 143 & 217 & 353 & 545 & 857\tabularnewline
\hline 
Bandwidth & $\Delta\nu/\nu$ & 30\% & 30\% & 30\% & 30\% & 30\% & 30\%\tabularnewline
\hline 
Number of spider web bolometers &  & 0 & 4 & 4 & 4 & 4 & 4\tabularnewline
\hline 
Number of polarization sensitive bolometers &  & 8 & 8 & 8 & 8 & 0 & 0\tabularnewline
\hline 
Angular resolution FWHM & arc mins & 9.5 & 7.1 & 5 & 5 & 5 & 5\tabularnewline
\hline 
Detector NET & $\mu K{s}^{1/2}$ & 50 & 62 & 91 & 277 & 1998 & 91000\tabularnewline
\hline 
$\Delta T/T$ Intensity (Stokes I) & $\mu K/K$ & 2.5 & 2.2 & 4.8 & 14.7 & 147 & 6700\tabularnewline
\hline 
$\Delta T/T$ Polarization (Stokes Q and U) & $\mu K/K$ & 4.0 & 4.2 & 9.8 & 29.8 &  & \tabularnewline
\hline
\end{tabular}

\caption{Planck HFI performance goals - Source:\citet{esa:2005}}
\label{tab:Planck_HFI}
\end{table}

From Appendix \ref{sec:CMB-Photon-Noise} we see that $NEP\simeq h\nu\sqrt{N}$,
where $N$ is the photon count. However, this is based on the assumption
inherent in the standard Big Bang model: that CMB photons lose energy
as they are red-shifted by the expansion of the universe. In the QSU,
no such expansion takes place in the cosmological reference frame
inhabited by these photons, and hence they can not be said to lose
energy. We therefore need to apply a redshift modification factor
to the apparent photon energy to give the correct value in the present
epoch, such that $E=h\nu(z_{0}+1),$ where $z_{0}$ is the redshift
at the time that the CMB photons were in blackbody equilibrium with
their surroundings, i.e. the era of recombination. Now if we apply
this corrected photon energy to the equation for the CMB energy density
(Equation \ref{eqn:dU}), we see that the predicted photon number
density (Equation \ref{eqn:RhoGamma}) will be reduced proportionately,
such that $N=N_{0}/(z_{0}+1)$. So the CMB photon noise level, using
the QSU model, will be given by:%
\footnote{It is worth noting that the same result can be obtained by assuming
that the CMB is emitted from a body at about $4800K$, but the integration
time in the reference frame of the CMB photons is reduced by a factor
of $1/\left(z_{0}+1\right)$.%
}

\[
NEP\simeq h\nu\sqrt{N(z_{0}+1)}\]

Applying this now to Equation \ref{eq:d/dT} gives us:

\begin{equation}
\frac{\Delta T}{T}=\frac{\left(e^{x}-1\right)^{2}}{xe^{x}}\sqrt{\frac{(z_{0}+1)(n^{2}+n)}{\Delta\nu\tau}}\end{equation}

This is plotted as the upper curve of Figure \ref{fig:photon_noise},
assuming a redshift at LSS of $z\simeq1800$. From this, it can be
seen that the predicted CMB photon noise level for a single detector
now lies well above the sensitivities of the Planck HFI bolometers.
In principle, therefore, it should be possible to measure this photon
noise in order to act as a test of the QSU model.

\section{Measurement strategy}

The scanning strategy employed on the Planck mission (see \citet{esa:2005})
is ideally suited to the generation of a data set that can be readily
analyzed to determine the level of photon noise in the CMB. The use
of multiple detectors in each HFI frequency band further extends the
options for analysis. The Planck satellite spins at one revolution
per minute, which with a beam size of 5 arc mins (FBHW), gives an
integration period per pixel per scan of approximately $14\times10^{-3}s$.
The satellite is re-pointed by an angle of 2.5 arc mins once every
hour (which provides two complete sky scans over a one year period),
so that each pixel will be scanned 60 times between re-pointings,
giving a total integration time of about $0.8s$. The data analysis
is somewhat complicated by the fact that the 2.5 arc min re-pointing
angle is designed to give a beam overlap between successive re-pointings
such that each sky pixel is sampled approximately 2.4 times. Various
alternative analysis strategies present themselves. For example, the
mean $\Delta T/T$ value for each detector chain (or pair of orthogonal
detectors in the case of polarization sensitive bolometers) could
be determined for each of the 4320 pixels in a 60 min sky scan. The
RMS deviation of each detector output from the mean value of all the
detectors would then provide a measure of the photon noise level. 

Since the precise orientation of the satellite in space is not critical
for the purposes of analyzing CMB photon noise, it should in principle
be possible to obtain the necessary data for this analysis from the
test measurements that will be carried out during the 3-4 months that
it will take the Plank satellite to reach its final orbital position
at the L2 point of the Earth-Sun system.

\section{Conclusions}

The Planck CMB mission will provide a wealth of detailed measurement
data that will help to pin down the value of various cosmological
parameters to hitherto unachievable levels of precision. Intriguingly,
the sensitivity of its detectors provides us with an ideal opportunity
to determine whether some of the fundamental assumptions underlying
the standard Big Bang model are in fact correct, or whether an alternative
paradigm might provide a more accurate description of our universe.

\pagebreak{}

\appendix{}

\section{Black Body Radiation}

Planck's black-body distribution law can be derived in two steps.
First, by considering the number of radiation modes that can be supported
in a black-body cavity of unit volume, an expression for the mode
density can be obtained

\begin{equation}
dN(\nu)=\frac{{8\pi\nu^{2}}}{{c^{3}}}d\nu\label{eqn:dN1}\end{equation}

The energy density as a function of frequency can then be calculated
by multiplying the mode density by the average energy per mode. In
classical terms this would simply be $kT$, where $k$ is the Boltzmann
factor and $T$ is the absolute temperature. However, the quantum
nature of photons results in the probability distribution being skewed,
so that the correct mean energy per mode must be calculated from the
Boltzmann distribution, giving

\begin{equation}
\bar{E}=\frac{h\nu}{e^{\nicefrac{h\nu}{kT}}-1}\label{eqn:Ebar}\end{equation}

Combining (\ref{eqn:dN1} and (\ref{eqn:Ebar}) we obtain the final
form of Planck's law for the energy density of black-body radiation
as a function of frequency

\begin{equation}
dU(\nu)=\frac{{8\pi h\nu^{3}}}{{c^{3}}}.\frac{d\nu}{e^{\nicefrac{h\nu}{kT}}-1}\label{eqn:dU}\end{equation}

The total energy density per unit volume is then simply obtained by
integrating (\ref{eqn:dU}) to give \begin{equation}
U=\frac{{8\pi^{5}k^{4}T^{4}}}{{15c^{3}h^{3}}}\label{eqn:U}\end{equation}

The final expression for the photon number density as a function of
frequency is merely (\ref{eqn:dU}) divided by the photon energy,
$h\nu$

\begin{equation}
dN(\nu)=\frac{{8\pi\nu^{2}}}{{c^{3}}}.\frac{d\nu}{e^{\nicefrac{h\nu}{kT}}-1}\label{eqn:dN2}\end{equation}

The photon number density per unit volume is therefore given by

\begin{eqnarray}
N & = & \int_{0}^{\infty}{\frac{{8\pi\nu^{2}}}{{c^{3}}}.\frac{d\nu}{e^{\nicefrac{h\nu}{kT}}-1}}\label{eqn:RhoGamma}\\
 & = & \frac{{16\pi k^{3}T^{3}\zeta(3)}}{{c^{3}h^{3}}}\end{eqnarray}

The energy intensity as a function of frequency and temperature is
given by

\[
I\left(\nu,T\right)=\frac{U(\nu,T)c}{4\pi}\]

Applying this to (\ref{eqn:dU}) yields

\begin{equation}
I\left(\nu,T\right)=\frac{2h\nu^{3}}{c^{2}}\frac{1}{e^{\nicefrac{h\nu}{kT}}-1}\label{eq:Intensity}\end{equation}

\section{CMB Photon Noise\label{sec:CMB-Photon-Noise}}

\subsection{Noise Equivalent Power}

Noise Equivalent Power (NEP) is defined as the incident signal power
required to obtain a signal-to-noise ratio of unity in a $1Hz$ bandwidth.
This derivation follows the methodology used in \citet{ben:1998}.

For blackbody radiation, the photon occupation number per mode is
given by:

\[
n=\frac{1}{e^{h\nu/kT}-1}\]

and the variance in occupation number is: \[
\langle\left(\Delta n\right)^{2}\rangle=n(1+n)\]

For a practical detector implementation, the mean square fluctuation
of photons \emph{detected} per mode is given by:

\[
\langle\left(\Delta n\right)^{2}\rangle=n(1+\eta\left(\nu\right)n)\]

where $\eta\left(\nu\right)$is the overall detector efficiency. 

The spectral density of the detected power is obtained by multiplying
this expression by the number of photon modes, $N$, and the energy
per photon, $h\nu$. 

\[
NEP_{v}^{2}=\frac{2N\eta\left(\nu\right)(h\nu)^{2}}{\left(e^{x}-1\right)}\left(1+\frac{\eta\left(\nu\right)}{(e^{x}-1)}\right)\]

where we have made the substitution $x=h\nu/kT$. 

To obtain the the total NEP we must integrate this expression over
the detector frequency band:

\[
NEP^{2}=\frac{1}{\tau}\int\frac{2N\eta\left(\nu\right)(h\nu)^{2}d\nu}{\left(e^{x}-1\right)}\left(1+\frac{\eta\left(\nu\right)}{(e^{x}-1)}\right)\]

where $\tau$ is the integration time of the detector measurement.

The number of photon modes is given by $N=\frac{A\Omega}{\lambda^{2}}$,
where $A$ is the area of the telescope and $\Omega$ is the solid
angle of the detector beam. In the diffraction limited case, $\lambda\simeq\sqrt{A\Omega}$
so $N\simeq1$. For a narrowband filter, where $\Delta\nu\ll\nu$,
we can approximate integrals so that this expression becomes:

\[
NEP^{2}=\frac{\eta(h\nu)^{2}\Delta\nu}{\tau\left(e^{x}-1\right)}\left(1+\frac{\eta}{(e^{x}-1)}\right)\]

giving us a final formula for the CMB photon noise power:

\begin{equation}
NEP=h\nu\sqrt{\frac{\Delta\nu(n^{2}+n)}{\tau}}\label{eq:NEP}\end{equation}

where $n=\eta/(e^{\nicefrac{h\nu}{kT}}-1)$.

\begin{figure}[h]
\includegraphics[bb=50bp 420bp 550bp 780bp,clip,scale=0.8]{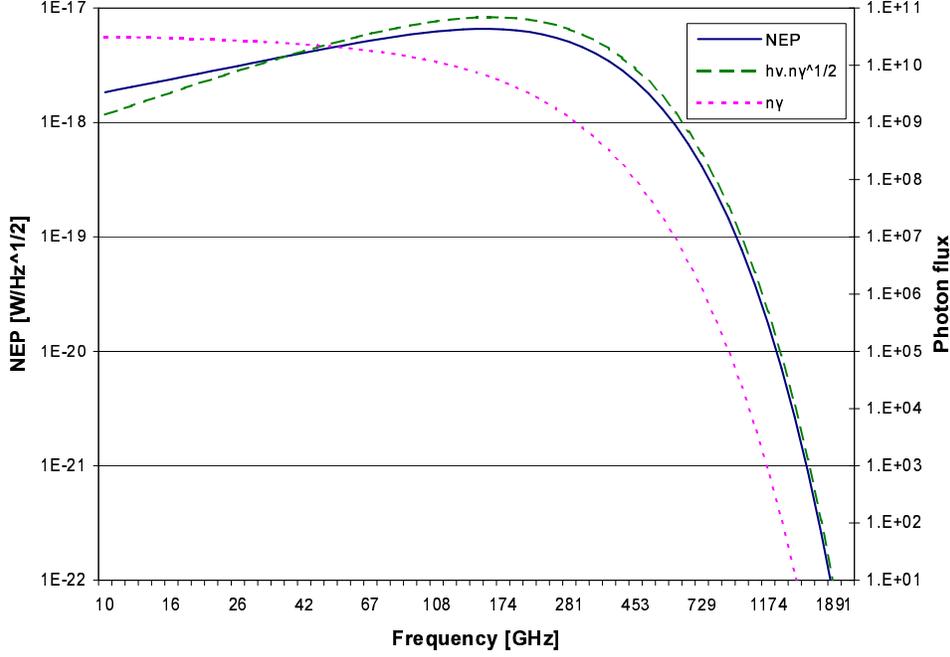}

\caption{NEP and CMB photon flux as a function of frequency}
\label{fig:NEP}
\end{figure}

Applying the appropriate values for the Planck satellite: $T_{CMB}=2.726$K,
frequency = 100 GHz, bandwidth = 30\%, detector efficiency $\eta=60\%$
, gives a $NEP\simeq1\times10^{-17}W/\sqrt{Hz}$. NEP and CMB photon
flux as a function of frequency are plotted in Figure \ref{fig:NEP}.
For frequencies higher than the Rayleigh-Jeans limit, given by $\nu=kT_{CMB}/h$,
it can be seen that the approximation $NEP\simeq h\nu\sqrt{n_{\gamma}}$
is valid.

\subsection{Thermodynamic units}

In the field of CMB astronomy, it is conventional to use thermodynamic
units in which variations in CMB brightness are expressed as temperature
variations with respect to the CMB background level. To derive an
expression for photon noise in thermodynamic units we start by differentiating
the equation for the mean energy per mode (\ref{eqn:Ebar}): 

\[
\bar{E}=\frac{h\nu}{e^{\nicefrac{h\nu}{kT}}-1}\]

Making the substitution $x=h\nu/kT$, and differentiating with respect
to $T$ gives:

\[
\frac{dE}{dT}=\frac{d}{dx}\frac{h\nu}{e^{x}-1}.\frac{d}{dT}x\]

\[
=\frac{h\nu e^{x}}{\left(e^{x}-1\right)^{2}}\frac{x}{T}\]

Applying this factor to the equation for NEP (\ref{eq:NEP}) we obtain
an expression for CMB photon noise in thermodynamic units:

\begin{equation}
\frac{\Delta T}{T}=\frac{\left(e^{x}-1\right)^{2}}{xe^{x}}\sqrt{\frac{n^{2}+n}{\Delta\nu\tau}}\label{eq:d/dT}\end{equation}

where $\Delta\nu$ is the bandwidth of the observation, $x=h\nu/kT$
is the dimensionless frequency, $\tau$ is the integration time, $n=\eta/\left(e^{x}-1\right)$
is the absorbed photon occupation number and $\eta$ is the absorption
efficiency. 

\pagebreak{}

\end{document}